# Tailing Magnetoelectric properties of $Cr_2Ge_2Te_6$ by Engineering Covalently bonded Cr Self-intercalation: Ferromagnetic Half-metal

Zhaoyong Guan,[†§*] Linhui Lv,[§] Ziyuan An,[§] Yanyan Jiang,[‡] Ya Su[ξ]

[†]Key Laboratory of Colloid and Interface Chemistry, Ministry of Education, School of Chemistry and Chemical Engineering, Shandong University, Jinan, Shandong 250100, P. R. China

[§]School of Chemistry and Chemical Engineering, Shandong University, Jinan 250100, P. R. China

[ξ] School of Electrical Engineering, Shandong University, Jinan, Shandong 250100, P. R. China

[‡]Key Laboratory for Liquid-Solid Structural Evolution & Processing of Materials (Ministry of Education), School of Materials Science and Engineering, Shandong University, Jinan, Shandong, 250061, People's Republic of China

[#]Research Center of Laser Fusion, China Academy of Engineering Physics. Mianyang, Sichuan 621900, P. R. China

# ABSTRACT

Two-dimensional intrinsic ferromagnetic half-metal (HM) are important for the spintronics. Manipulating the interlayer magnetic coupling of van der Waals magnetic materials is an important method to control magnetoelectric properties, which is especially useful for the spintronics. Here, based on systematical research of $CrGeTe_3$ (CGT) bilayer and multilayers with s of self-intercalated (SI) Cr atom, we find that self-intercalation can enhance the interlayer magnetic coupling. The super-exchange interaction still dominates interlayer magnetic exchange interaction, which results in ferromagnetic coupling between neighboring vdW layers. CGT bilayer keeps HM with FM order after Cr self-intercalation, independent of self-intercalated Cr ($Cr_{SI}$) atoms' concentration. Most importantly, self-intercalated $CrGeTe_3$ (SI-CGT) bilayers show ferromagnetic HM, independent of stacking orders. Moreover, SI-CGT multilayers keep FM order, independent of films' thickness. However, SI-CGT multilayers transform from HM into normal spin-polarized metal, as the states at the Fermi-level increases. Moreover, magnetic anisotropy energy (MAE) of SI-CGT-AA and SI-CGT-AB are -0.160 and -0.42 meV/.f.u., which are modulated by $Cr_{SI}$ atoms. The MAE of SI-CGT-AA and SI-CGT-AB are different, as the hybridization interaction between Cr's $d$ orbitals is different. SI-CGT multilayers' magnetic easy axis (EA) switches from [001] of CGT to [100] direction, independent of stacking orders. It origins that MAE mainly contributed by hybridization between Te's $p_x$ and $p_y$, $p_y$ and $p_z$ orbitals is obvious weakened as $Cr_{SI}$ is introduced. SI-CGT multilayers show good dynamical, thermal, and magnetic stability at 300 and 500 K. These findings find a promising way to manipulate interlayer exchange interaction and magnetoelectric properties of CGT multilayers and other vdW magnets.

## 1. INTRODUCTION

Since discovery of graphene, all kinds of two-dimensional (2D) materials, such as hexagonal boron-nitride,[1, 2] transition metal dichalcogenides,[3, 4] and stanene[5] have been found. Most 2D materials are non-magnetic materials,[6, 7] as long-range magnetic order is prohibited by the strong fluctuations at finite temperature in 2D materials.[8] Recently, 2D van der Waals (vdW) magnetic materials,[9, 10] such as $CrI_3$ ($MI_3$),[11-17] $Fe_3GeTe_2$,[18, 19] $VS(Se)_2$,[20-22] $VSeTe$,[23, 24] and CGT[25-27] have attracted much attention. Interlayer magnetic coupling has been confirmed to play an important role in modulating magnetoelectric properties of vdW magnetic materials. Taking $CrI_3$ as an example, there is two stacking patterns, which show interlayer ferromagnetic (FM) and antiferromagnetic (AFM) coupling, respectively.[28]

In all kinds of modification methods, intercalation is a useful approach for controlling vdW materials' properties,[29] such as topological properties,[30] photonics,[31] optoelectronic properties,[32, 33] thermoelectric properties,[34, 35] energy storage,[36, 37] catalytic properties,[38] and magnetoelectric properties.[16, 17, 27, 39-47] Apart from foreign atoms as intercalants, the native atoms could also work as interaction.[47] It's named self-intercalation, which attracts much attention. The SI structures show more stable structure, compared with other vdW structures.[29] In addition, the native SI magnetic atoms could affect interlayer magnetic coupling.[39, 41, 48, 49] In 2020, the researchers successfully synthesized SI structures in TaS(Se)$_x$ by performing growth under high metal chemical potential.[47, 48] Besides that, $V_{11}S_{16}$, $In_{11}Se_{16}$ and $Fe_xTe_y$ could be synthesized under metal-rich condition.[47] In 2022, the phase-selective synthesis of trigonal and monoclinic $Cr_5Te_8$ crystals have been achieved, using chemical vapor deposition.[50] The successfully experimental SI realization in 2D materials,[51] which paves a new way of modulating magnetoelectric properties of 2D materials.[47, 48, 50]

In this article, by using first-principles method, we perform systematic research on magnetic and electronic properties of SI-CGT bilayers and multilayers. The SI-CGT bilayers are HM with FM order, independent of stacking patterns and $Cr_{SI}$ atoms' concentration. SI-CGT multilayers are spin-polarized metal with FM order, regarding of stacking patterns and films' thickness. The AB stacking orders has lower energy when the number of layers is less than five. Moreover, $Cr_{SI}$ atoms could effectively modulate MAE by switching EA from [001] to [001] direction, as the hybridization between Te's $p_x$ and $p_y$, $p_y$ and $p_z$ orbitals is obviously weakened. The different stacking orders affect the hybridization between Cr atoms' $d_{xz}$ and $d_{xy}$, $d_{xy}$ and $d_{x^2-y^2}$ orbitals, resulting in the different MAE of SI-AA and SI-AB stackings. SI-CGT multilayers show good dynamical, thermal, and magnetic stability. These findings find a promising way to manipulate interlayer exchange interaction and magnetoelectric properties of CGT multilayers and other vdW magnets.

## 2. COMPUTATIONAL DETAIL

The calculation of SI-CGT is using plane-wave basis Vienna Ab initio Simulation Package (VASP) code,[52] based on the density functional theory. The generalized gradient approximation (GGA) with Perdew-Burke-Ernzerhof (PBE)[53] is adopted. GGA+U method[54] and hybrid-functional HSE06[55, 56] are used to deal with strong-correlated correction to Cr's 3$d$ electrons. The effective onsite Coulomb interaction parameter ($U$) and exchange interaction parameter ($J$) are set to be 3.60 and 0.60 eV, respectively. Therefore, the effective $U$ ($U_{eff}$, $U_{eff} = U - J$) is set 3.0 eV.[57, 58] And MAEs change with $U_{eff}$ is also tested, shown in Figure S1. The geometry optimization, energies with all kinds of magnetic orders, band structure, density of states (DOS), phonon spectrum, *ab* initio molecular dynamics (AIMD), and MAE is calculated with LDA+U method. Energies with all kinds of magnetic orders, and DOS are confirmed by HSE06 functional. The vacuum space in the *z*-direction is set 16 Å to avoid virtual interaction. The kinetic energy cutoff is set as 360 eV. The geometries are fully relaxed until energy and force is converged to $10^{-5}$ eV and 1 meV/Å, respectively. 9×9×1 and 12×12×1 Monkhorst-Pack grids[59] are used for geometry optimization and energy calculation, respectively. Spin-orbital coupling (SOC) is also considered in the calculation.

MAE is calculated with an energy cutoff of 400 eV. The convergence of total energy and force are $1\times10^{-8}$ eV and $1\times10^{-3}$ eV/Å, respectively. Based on systematical test, the corresponding *k*-grid is adopted 21×21×1, without any symmetry constriction, shown in Figure S2. Phonon spectra and DOS are calculated using finite displacement method as implemented in Phonopy software.[60] A 2×2×1 cell is adopted in the simulation, and total energy and Hellmann-Feynman force are converged to $10^{-8}$ eV and 1 meV/Å, respectively. 6000 uniform *k*-points along high-symmetry lines are used to obtain phonon spectra. Moreover, AIMD simulation is also performed to confirm structural and magnetic stability. The constant moles–volume–temperature (NVT) ensemble with Nosé–Hoover thermostat[61] is adopted at temperature of 300 and 500 K, respectively. The time step and total time are 1 fs and 10 ps, respectively. In order to eliminate effect of periodic boundary condition with relatively smaller system size, a larger supercell (2×2×1 cell) is used in AIMD simulation.

DFT-D2 Grimmer method[62] is adopt to describe weak vdW interaction between layers, and DFT-D3 Grimmer[63] is also used to confirm key results.

The climbing image-nudged elastic band (CI-NEB) method[64] is used to calculate migration barrier of $Cr_{SI}$ atom. And atomic positions are fully relaxed with a force criterion of 0.03 eV/Å$^{-1}$ for all intermediates and transition states.

## 3. RESULTS AND DISCUSSION

**3.1. Geometry of SI-CGT Monolayer (ML).** There are two different stacking orders: AA and AB for CGT bilayer. Therefore, when $Cr_{SI}$ atoms are embedded, there are two kinds of structures, named AA-SI and AB-SI, shown in Figure 1 a-f. The geometries of AA-SI and AB-SI are fully optimized by LDA+U method. 111-AA-SI bilayer (111-AA-SI-2) shows $D_{3D}$ point group, while 111-AB-SI bilayer (111-AB-SI-2) shows $C_3$ point group, shown in Figure 1 a-c, and d-f, respectively. Both 111-AA-SI-2 and 111-AB-SI-2 have same lattice parameter ($a=b$) of 6.913 Å, shown in Figure 1a, d. The lattice parameter of AA-SI and AB-SI are similar with CGT of 6.910 Å.[10] The vertical distance between Te atoms of different layers is 3.896 Å (111-AA-SI-2), and 3.816 Å (111-AB-SI-2), respectively. These values are larger than CGT (bulk) of 3.4 Å,[10] as Cr atoms are inserted. The distance between $Cr_{SI}$ and nearby Te atoms is 2.913 and 2.910 Å, respectively. The $Cr_{SI}$ atom intends to locate at hollow site of top and bottom layers, shown in Figure 1. The corresponding angels between $Cr_{SI}$ and Te atoms in111-AA-SI-2 are 105.59°, 94.79°, and 80.16°, while these angels are 81.41°, 98.59°, 98.59° for 111-AB-SI-2, respectively. The intralayer of AA-SI-2 and AB-SI-2 keeps original structure, without obvious distortion, shown in Figure 1 a-f. The geometries of 11-AA-SI-2 and 111-AB-SI-2 are also optimized by DFT-D3 method, shown in Figure S3 a, b, respectively, and distance between $Cr_{SI}$ and nearby Te atoms is little smaller than DFT-D2 method. More detail could be found in supporting information.

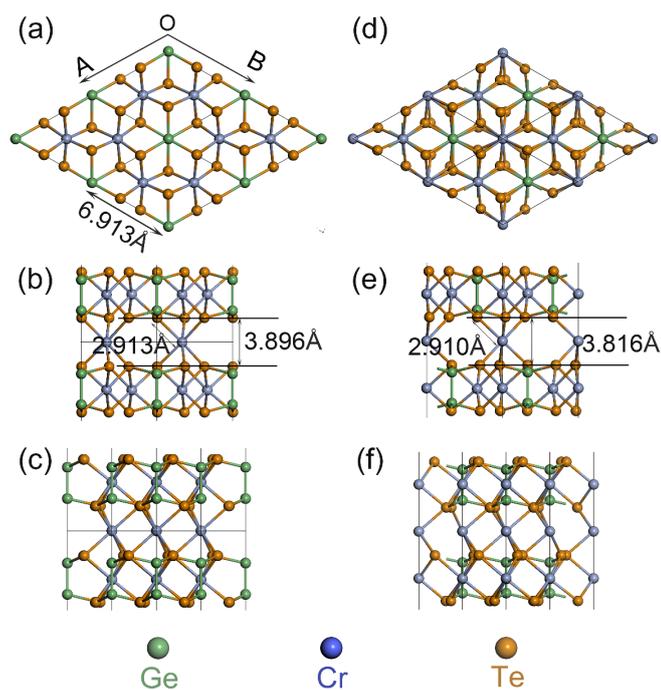

**Figure 1.** (a, d) Top, (b, e) side-1 and (c, f) side-2 views of optimized geometry of 111-AA-SI-2 and 111-AB-SI-2. The green, blue and yellow balls represent Ge, Cr, and Te atoms, respectively.

## 3.2. Magnetic and Electronic Properties.

The geometries of SI-CGT are investigated, and magnetic properties are usually related with geometries. In this section, the magnetic properties are studied. The magnetic moment (MM) mainly localizes at Cr atoms, shown in Figure 2 a-f. In order to confirm magnetic ground state, eleven magnetic orders are considered, shown in Figure S4 a, b, respectively. It can be concluded that 111-AA-SI-2 and 111-AB-SI-2 show FM (FM-$_{FM-FM}$) order. The FM-$_{FM-FM}$ order is defined as followed: Cr atom of intralayer ferromagnetically couple with other Cr atoms, and Cr atom of interlayer ferromagnetically couple with other Cr atoms, shown in Figure 2 a, b, d, e. The AFM (AFM-$_{FM-FM}$) order is defined: Cr atom of intralayer ferromagnetically couple with other Cr atoms, while Cr atom of interlayer antiferromagnetically couple with other Cr atoms, shown in Figure 2 c, f. The corresponding energy difference ($\Delta E$) between FM and AFM orders is defined to describe magnetic stability of materials: $\Delta E = E_{FM} - E_{AFM}$. 111-AA-SI-2 and 111-AB-SI-2 with FM order are 0.266 and 0.278 eV lower than AFM-$_{FM-FM}$ order, which means they show FM order. And the corresponding $\Delta E$ are -0.266 and -0.278 eV, respectively. However, $\Delta E$ of CGT bilayer without Cr$_{SI}$ atoms are -2.680 (111-AA-2), and -4.371 meV (111-AB-2), respectively.[10] It can be concluded that SI could obviously enhance ferromagnetism of CGT, as super-exchange interaction between Cr atoms is strengthened. And differential charge density is defined as followed: $\Delta\rho = \rho_{\text{CGT-SI}} - \rho_{\text{CGT-1}^{\text{st}}\text{L}} - \rho_{\text{CGT-2}^{\text{nd}}\text{L}} - \rho_{\text{Cr-SI}}$. And charges accumulate and deplete at $\text{Cr}_{\text{SI}}$ atoms, by analyzing differential charge density, shown in Figure S3 a-d. It can be found $\text{Cr}_{\text{SI}}$ atom loses 0.78 (0.73) $e$ electron to the nearby Te atoms by the bader analysis.[65] The Te atoms (1$^{\text{st}}$ L) of 111-AA-SI-2 bonding to the $\text{Cr}_{\text{SI}}$ atom lose 1.075, 1.253 1.233 $e$ electrons, and Te atoms (2$^{\text{nd}}$ L) lose 1.228, 1.277, and 1.280 $e$ electrons. Te atoms (1$^{\text{st}}$ L) of 111-AB-SI-2 lose 1.626, 1.352, 1.267 $e$ electrons, and Te atoms (2$^{\text{nd}}$ L) lose 1.503, 1.206, and 1.274 $e$ electrons, respectively. More detail could be found in Supporting Information. The Cr atom has $d^5s^1$ configuration, and one $d$ electron and $s$ electron are lost. As a result, Cr atoms of 111-AA-SI-2 with FM order have 3.683 (2rd L), 3.683 (1$^{\text{st}}$ L), 3.911 (2rd L), 3.911 (1$^{\text{st}}$ L), 4.180 ($\text{Cr}_{\text{SI}}$) $\mu_B$ MM, respectively. For 111-AB-SI-2, Cr atoms have 3.683 (2rd L), 3.683 (1$^{\text{st}}$ L), 3.915 (2rd L), 3.915 (1$^{\text{st}}$ L), 4.160 ($\text{Cr}_{\text{SI}}$) $\mu_B$ MM, respectively. Moreover, more magnetic orders and corresponding energies are calculated, shown in Figure S4 a, b, respectively. Both 111-AA-SI-2 and 111-AB-SI-2 show FM ground state, in considering magnetic orders. In a word, the super-exchange interaction plays an important role in interlayer magnetic exchange interaction.

The electronic properties are usually related with magnetic orders, and band structures of 111-AA-SI-2 and 111-AB-SI-2 with FM order are calculated, shown in Figure 2 g, h, respectively. There are two sub-bands of 111-AA-SI-2 crossing Fermi-level for spin-α electrons. And these states at Fermi-level are mainly contributed by Cr and Te atoms, shown in Figure 2 g, h. Therefore, the spin-α electrons are conducting. However, 111-AA-SI-2 is semiconductor with an indirect band gap of 0.682 eV for spin-β electrons. In a word, 111-AA-SI-2 is HM. The similar phenomenon also appears in 111-AB-SI-2. The spin-α electrons of 111-AB-SI-2 are conducting, while spin-β electrons are semiconductive. Therefore, 111-AB-SI-2 is also HM. Other stacking orders are tested, and they are all HM with FM order, independent of stacking orders. In sum, both 111-AA-SI-2 and 111-AB-SI-2 are HM with FM order, and these states near the Fermi-level are mainly contributed by Cr atoms, shown in Figure 2 g, h, respectively.

In order to clarify Cr's magnetic moment, $d$ electrons' PDOS and integrated density of the states (IDOS) are calculated, shown in Figure 3. The corresponding Cr projected PDOS are shown in Figure 3 a, b, respectively. The $d_{x^2-y^2}$, $d_{z^2}$, $d_{yz}$, $d_{xz}$, and $d_{xy}$ orbitals of 111-AA-SI-2 are partially occupied by spin-α electrons, shown in Figure 3 a, c. Moreover, $d_{yz}$ and $d_{xz}$ orbitals of 111-AA-SI-2 are degenerate, shown in Figure 3 a, c. $d_{x^2-y^2}$ and $d_{xy}$ orbitals of 111-AA-SI-2 are also degenerate, shown in Figure 3 a, c. The similar trend also appears in 111-AB-SI-2, shown in Figure 3 b, d. For 111-AB-SI-2, $d_{yz}$ and $d_{xz}$ orbitals are degenerate, which make main contribution to *d*-orbital. In addition, $d_{x^2-y^2}$ and $d_{xy}$ orbitals are also degenerate, shown in Figure 3 b, d. In fact, other stacking orders are also tested, which show the similar trend. All $d_{x^2-y^2}$, $d_{xy}$, $d_{yz}$ and $d_{xz}$ orbitals are degenerate, respectively.

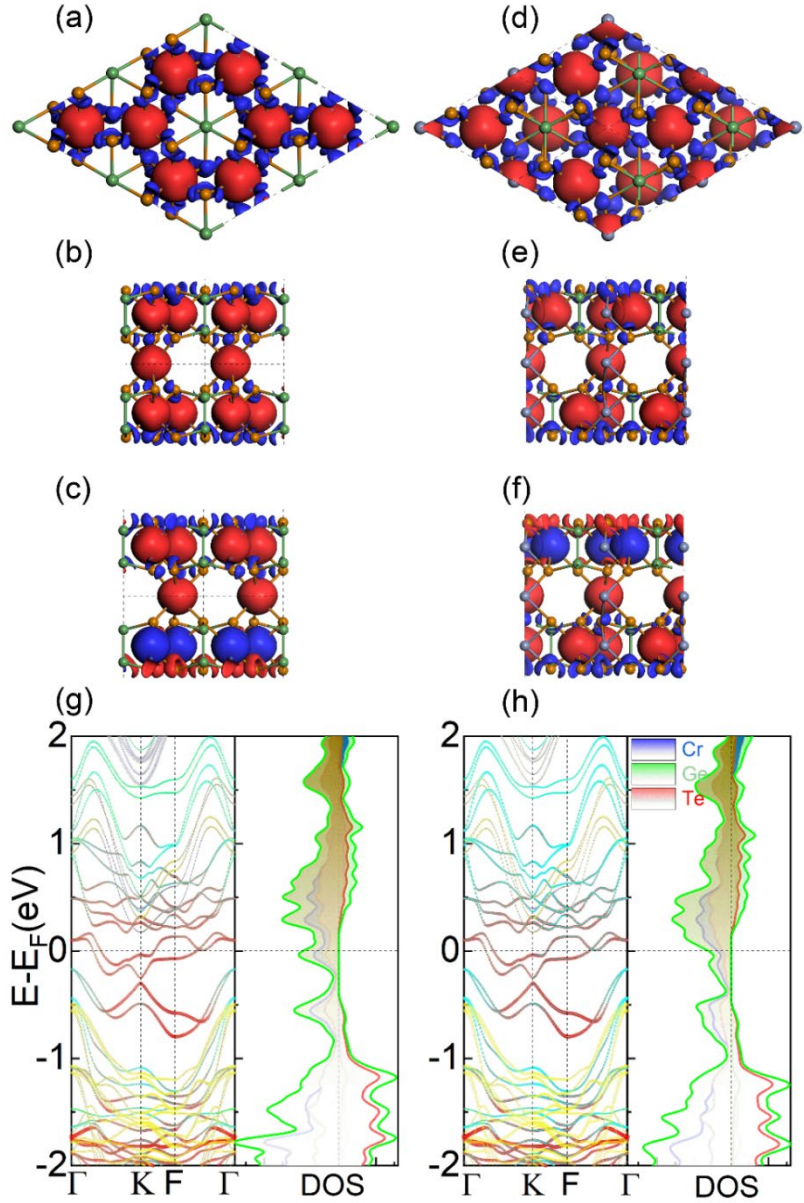

**Figure 2.** The spin charge density difference of 111-AA-SI-2 with (a, b) FM, (c) AFM orders, respectively. The spin charge difference density of 111-AB-SI-2 with (d, e) FM, and (f) AFM orders. The spin-polarized band structure and PDOS of (g) 111-AA-SI-2 and (h) 111-AB-SI-2, respectively. The red, gray, green, blue, cyan, and yellow represent Cr-α, Cr-β, Ge-α, Ge-β, Te-α and Te-β atoms projected band structure, respectively. The blue, green, red present Cr, Ge, Te atoms' PDOS of SI-CGT.

Moreover, the electronic properties are usually related with magnetic orders. Both 111-AA-SI-2 and 111-AB-SI-2 show FM order. And band structures with other magnetic orders are also calculated, shown in Figure S 5，6, respectively. 111-AA-SI-2 with FM-$_{FM-FM}$, AFM-$_{FM-FM}$ (Figure S5c), and +−−−− (1$^{st}$ L-SI-2$^{nd}$ L) (Figure S5g) orders are HM, while other orders are normal spin-polarized metal, shown in Figure S5. However, only 111-AB-SI-2 with FM-$_{FM-FM}$ order is HM, while other orders are normal spin-polarized metal, shown in Figure S6. More detail and discussion are shown in Figure S 5, 6, respectively.

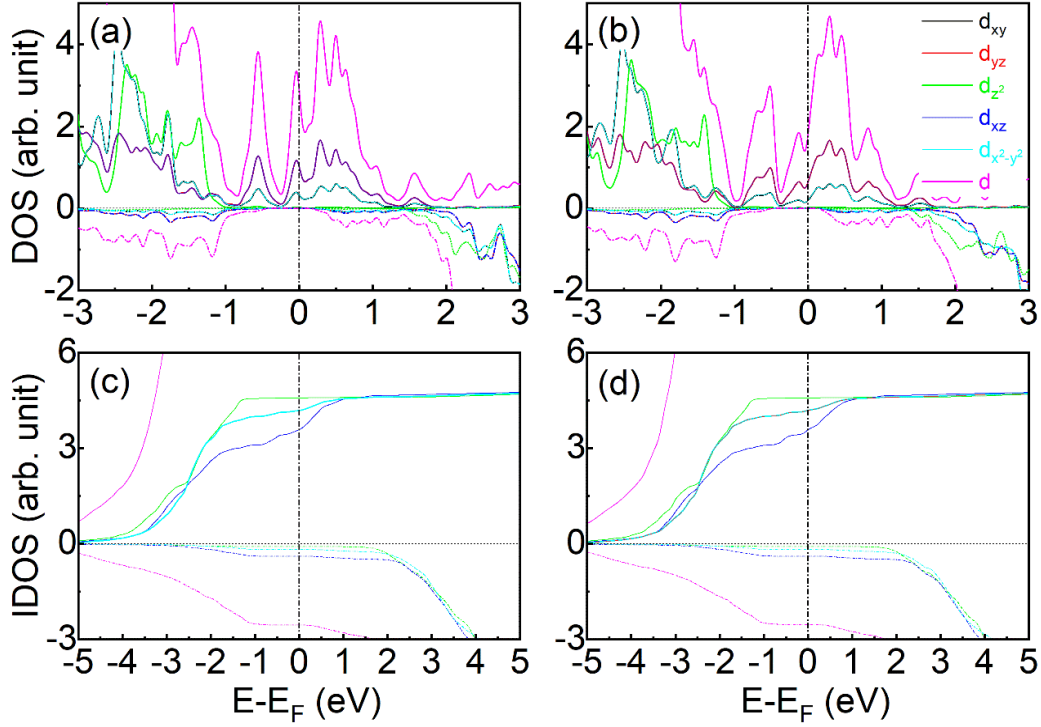

**Figure 3.** Co's *d*-orbital projected DOS of (a) 111-AA-SI-2, (b) 111-AB-SI-2, IDOS of (c) 111-AA-SI-2, and (d) 111-AB-SI-2. The black, red, green, blue, cyan and pink represent $d_{xy}$, $d_{yz}$, $d_{z^2}$, $d_{xz}$, $d_{x^2-y^2}$ and total $d$ orbitals projected DOS, respectively. The solid, and dashed lines represent spin-α and spin-β electrons, respectively.

In order to confirm validity of vdW corrected functionals, other vdW corrected functionals are also adopted. The results of DFT-D3 method are shown in Figure S7. The spin charge density difference of 111-AA-SI-2 and 111-AB-SI-2 are shown in Figure S7 a-b, which is the same with DFT-D2. Both 111-AA-SI-2 and 111-AB-SI-2 are HM with FM order, shown in Figure S7 c-d. More detail could be found in Figure S7, in the supporting information.

The art of HSE06 functional is also used to calculate spin charge density difference and PDOS of 111-AA(AB)-SI-2, shown in Figure S8 a-f. Both 111-AA-SI-2 and 111-AB-SI-2 show FM order. And the corresponding $\Delta E$ are -28.3, -30.1 meV, shown in Figure S8 a-d. 111-AB-SI-2 has lower energy than 111-AA-SI-2, which is consistent with PBE+U method. The corresponding PDOS are shown in Figure S8 e-f, and they are all HM, independent of stacking patterns. However, spin-β electrons' gap calculated by HSE06 is 1.238 eV, which is larger than PBE+U (0.682 eV). More detail could be found in the supporting information.

As there is Cr atom in CGT, the effect of SOC on electronic properties should consider. The band structure with SOC is also calculated, shown in Figure S9, in the supporting information. There is a Dirac cone at K point, as magnetic axis is along [100] direction. The band structures with SOC are metal, regard of stacking orders for SI-CGT bilayer.

**3.3. Properties of Different Concentration of $Cr_{SI}$ Atoms.** In order to investigate electronic properties of different $Cr_{SI}$ atoms' concentration, larger N×N×1 ( N = $\sqrt{3}$, 2, $\sqrt{7}$, 3, 4, 5, 6 ) supercells are fabricated. The corresponding $Cr_{SI}$ atoms' concentration (*x*) is 33.33 %, 25 %, 14.29%, 11.11%, 6.25%, 4%, and 2.78%, whose $\Delta E$ of NN1-AA-SI-2 is -0.285, -0.223, -0.267, -0.275, -0.159, -0.128, and -0.176 eV, respectively, shown in Figure 4a, S10. For NN1-AB-SI-2, the corresponding $\Delta E$ is -0.296, - 0.245, -0.288, -0.276, -0.243, -0.152, and -0.286 eV, respectively. $\Delta E$s are always negative, which implies NN1-AA-SI-2 and NN1-AB-SI-2 show FM order, independent of *x*. Moreover, $\Delta E$ of NN1-AA-SI-2 are always smaller than NN1-AB-SI-2. And the energy difference ($\Delta E_{AA-AB-FM}$) between $E_{AA-FM}$ and $E_{AB-FM}$ is defined as followed:

$$\Delta E_{AA-AB-FM} = E_{AA-FM} - E_{AB-FM} \quad (1)$$

The $\Delta E_{AA-AB-FM}$ are 0.042, 0.366, 0.10 and 0.025 eV for $N = 1, \sqrt{3}, 2, \sqrt{7}$, respectively. As the *x* is further decreased, the corresponding $\Delta E_{AA-AB-FM}$ is 0.062, 0.09, 0.186, and 0.532 eV for $N = 3, 4, 5, 6$. The $\Delta E_{AA-AB-FM}$ is always positive, which implies NN1-AB-SI-2 are always more stable than the corresponding NN1-AA-SI-2, shown in Figure 4 b. And this trend also appears in the bulk of CGT.[10]

As *x* is decreased, the corresponding electronic properties also change, shown in Figure 4 c-l, respectively. The PDOS of 221-AA-SI-2 and 221-AB-SI-2 show DOS at Fermi-level ($DOS_F$) of 9.272, 6.950 states/eV, respectively. As *x* is further decreased, the corresponding $DOS_F$ are 11.559 (9.974), 12.256 (13.985) states/eV for $N = 3, 4$, shown in Figure 4 d, e, i, j, respectively. As *x* is decreased to 4.0%, 2.78%, the corresponding $DOS_F$ are 13.635 (16.091), 12.612 (17.450) states/eV, shown in Figure 4 f, g, k, l, respectively. In a word, all spin-α electrons are conducting, while all spin-β electrons are semiconductive. Therefore, all considering

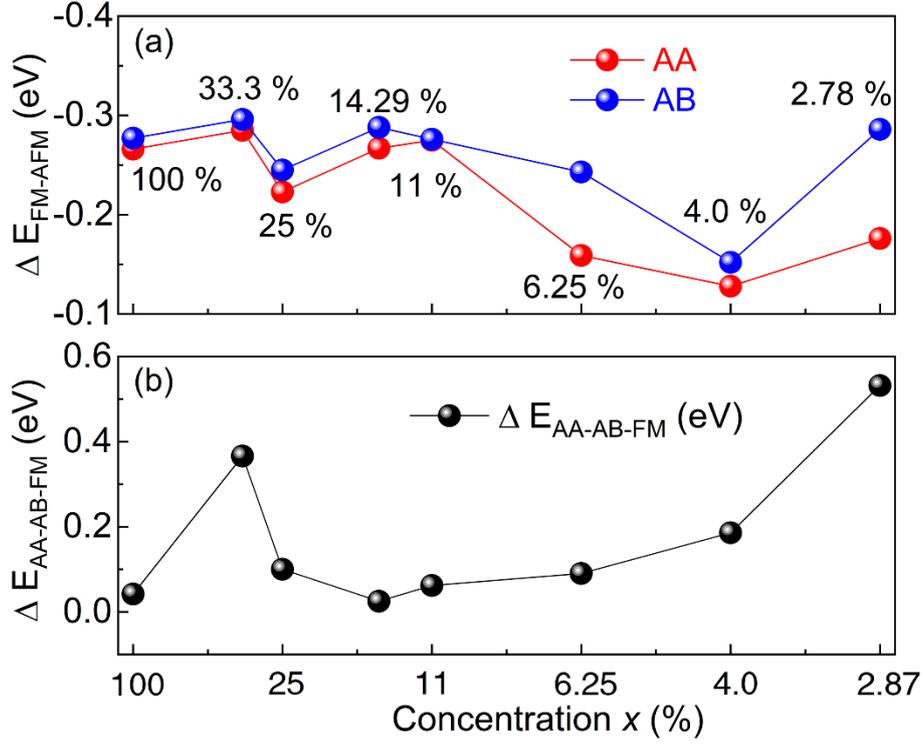

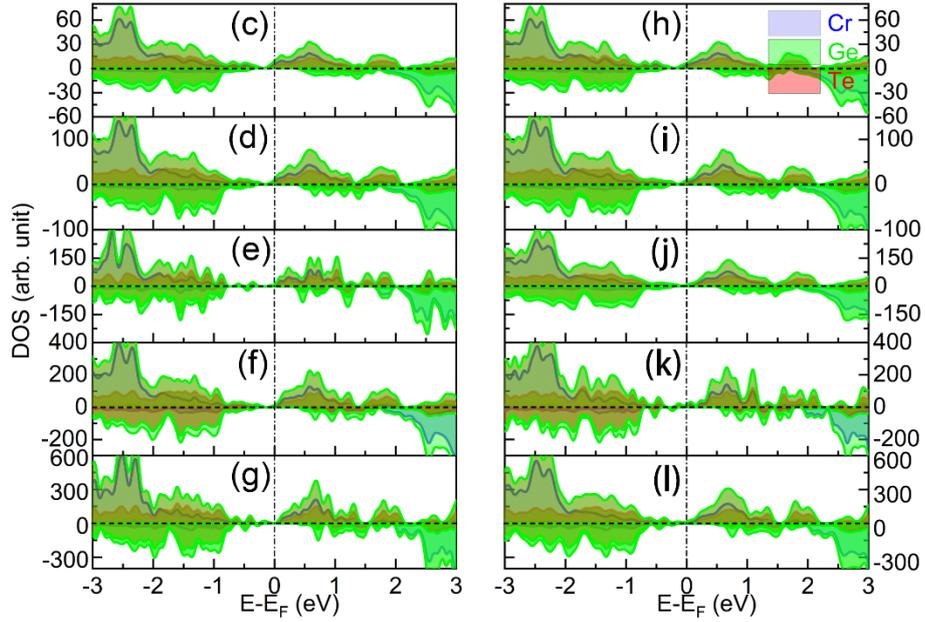

**Figure 4.** (a) $\Delta E_{FM-AFM}$ and (b) $\Delta E_{AA-AB-FM}$ change with concentration of $Cr_{SI}$ atoms. The PDOS of NN1-AA-SI-2 with N equals to (c) 2, (d) 3, (e) 4, (f) 5, and (g) 6. PDOS of NN1-AB-SI-2 with N equals to (h) 2, (i) 3, (j) 4, (k) 5, and (l) 6. The Fermi-level is set 0 eV.

geometries show HM with FM order, independent of $Cr_{SI}$ atoms' concentration.

**3.4. Magnetic Anisotropy Properties.** MAE is often used to describe magnetic stability of materials, and a larger MAE is highly expected in the HM. The larger MAE means electrons need more energy to overcome a "higher" barrier from EA to hard axis. In this part, MAE of CGT and SI-CGT is calculated using LDA+U with SOC. The corresponding energy ($\Delta E_0$) along certain direction ($\theta$, $\phi$) follows these equations:

$$\Delta E_0 = K_1 \cos^2\theta + K_2 \cos^4\theta + + K_3 \cos^6\theta + K_4 \cos 3\phi \quad (2)$$
$$\Delta E_0 = E - E_{[001]} \quad (3)$$

Where $E_{[001]}$ represents energy along magnetic axis of [001] direction. $K_1$, $K_2$ and $K_3$ stand for the quadratic, quartic and sextic (six degree) contribution to MAE, respectively. $\phi$ represents azimuthal angel, while $\theta$ represents polar angle. As 111-AA-2 and 111-AA-SI-2 have $D_3$ point group, while 111-AB-2 and 111-AB-SI-2 have $C_3$ point group. The energy difference $\Delta E_0$ is independent of in-plane azimuthal angel $\phi$.[16, 17, 27] And the energies with EA along [100] and [010] directions are the same.[27] As a result, $K_4$ equals to 0, shown in Figure S11 b, c, 5 a-b, respectively. And equation-3 is also simplified into following equation:

$$\Delta E_0 = K_1 \cos^2\theta + K_2 \cos^4\theta + K_3 \cos^6\theta \quad (4)$$

$\Delta E_0$ changes with functions $\phi$ and $\theta$, shown in Figure 5 a, b, S11 b, c, respectively. For 111-AA-2 and 111-AB-2, $\Delta E_0$ follows these equations: $\Delta E_0 \text{ (meV)} = 1.79\cos^2\theta + 0.010\cos^4\theta$, $\Delta E_0 \text{ (meV)} = 1.89\cos^2\theta + 0.016\cos^4\theta$, shown in Fig S11b. Compared with $k_2$, $k_1$ is much larger, which means quadratic part makes main contribution to MAE. When $Cr_{SI}$ atoms are embedded into CGT bilayers, $\Delta E_0$ of 111-AA-SI-2 follows this equation: $\Delta E_0 \text{ (10}^{-1}\text{meV)} = -1.834\cos^2\theta + 0.603\cos^4\theta - 0.356\cos^6\theta$. As 111-AB-SI-2, $\Delta E_0$ follows this equation: $\Delta E_0 \text{ (10}^{-1}\text{meV)} = -4.019\cos^2\theta - 0.308\cos^4\theta - 0.047\cos^6\theta$, shown in Figure 5b. $k_1$ are -0.183, and -0.4019, while $k_2$ are 0.0603, and -0.0308, respectively. Compared with $k_2$ and $k_3$, $k_1$ is much larger, which is similar with CGT, shown in Figure S11b. According to Figure 5b, MAE could be calculated by the following equations:[27]

$$MAE = E_{[100]} - E_{[001]} \qquad (5)$$

The corresponding MAEs of 111-AA-2 and 111-AB-2 are 1.80, and 1.91 meV/f.u., shown in Figure S11c, respectively. The positive MAE implies EA is along out-plain direction, that is, parallel to the c-axis, which intends to perpendicular magnetic anisotropy (PMA), shown in Figure S11a.[10] When $Cr_{SI}$ atoms are inserted, the corresponding MAEs of 111-AA-SI-2 and 111-AB-SI-2 are -0.160 and -0.428 meV/.f.u., respectively, shown in Figure 5d. The negative MAE implies EA is along in-plain direction, which corresponds to in-plane anisotropy (IPA), shown in Figure 5c. The MAE transforms from the positive to the negative, which means EA is switched from the out-plane to the in-plane, when the $Cr_{SI}$ atoms are introduced.

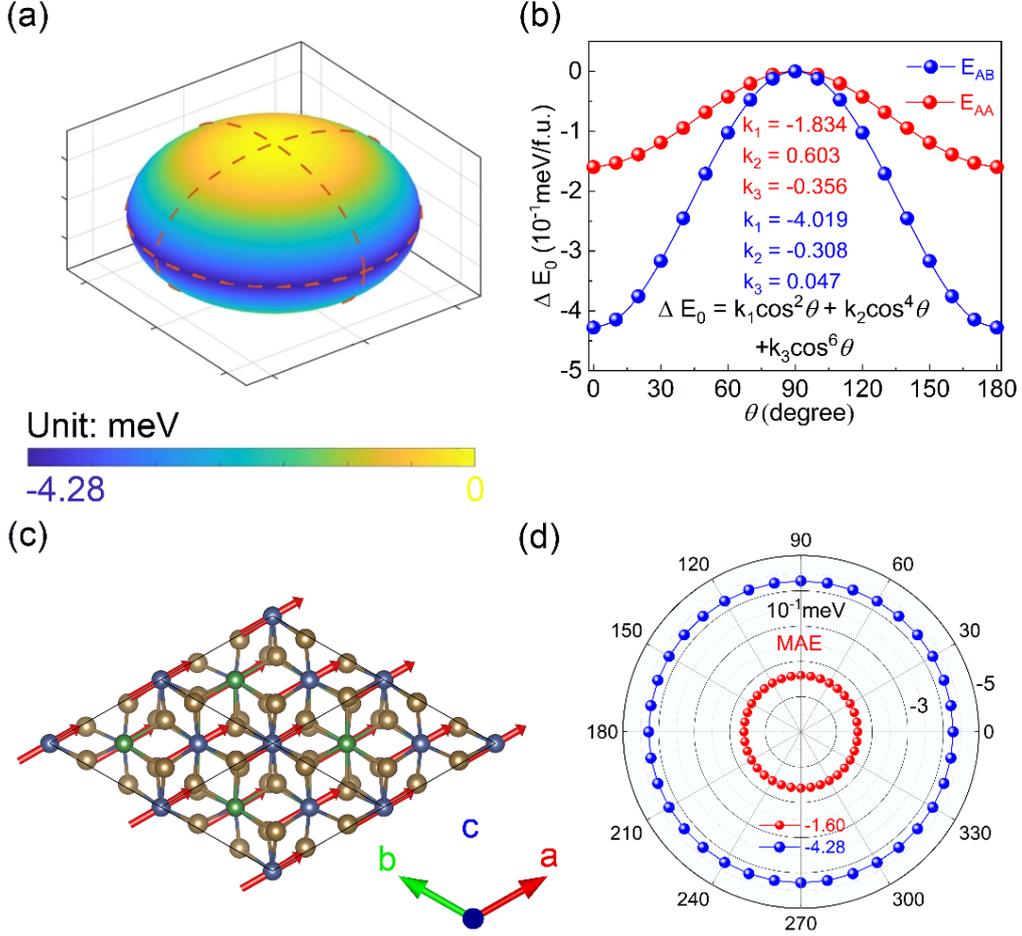

**Figure 5**. The MAE map (FM state as a reference with EA along [100]) of (a)111-AA-SI-2. The energy indicated by the dashed lines changes with azimuthal angle $\varphi$. (b) $\Delta E_0$ varies from the out-of-plane to the in-plane direction of 111-AA-SI-2 and 111-AB-SI-2. (c) The red arrow represents direction of EA (along [100] direction) of 111-AA-SI-2 bilayer. (d) $\Delta E_0$ changes with polar angle $\theta$. Red, and blue lines represent 111-AA-SI-2 and 111-AB-SI-2, respectively.

MAE of SI-CGT is related with the stacking orders, and MAE of 111-AB-SI-2 is larger than 111-AA-SI-2. And it's caused by different hybridization interaction between Cr's orbitals of stacking patterns. Additionally, EA of 111-AA-SI-2 and 111-AB-SI-2 intend to IPA, which is different from PMA of CGT bilayers. It's caused by the $Cr_{SI}$ atom, and more discussion could be found in other part.

**3.5. Multilayers with SI.** When CGT ML is stacked together, it could form all kinds of stacking patterns. In this section, magnetic and electronic properties of stacking multilayers with self-intercalated Cr atoms are systematically investigated. As layer number is increased to three, 111-AA-SI-3 and 111-AB-SI-3 are fabricated, and the corresponding $\Delta E$ is 0.49, and 0.51 eV, respectively. When layer numbers (N) are further increased to four, five, and six, the corresponding $\Delta E$ are 0.792 (AB 0.812), 1.003 (1.026), and 1.272 (1.362) eV, shown in Figure 6a. When $N = 7, 8, 9, 10$, the corresponding $\Delta E$ are 1.495 (1.570), 1.777 (1.914), 2.004 (2.099), and 2.215 (2.356) eV, respectively. Most interesting, $\Delta E_{AA}$ and $\Delta E_{AB}$ follow these equations: $\Delta E_{AA}= -0.223+0.247N$ and $\Delta E_{AB}= -0.258+0.264N$. As the film thickens, both $\Delta E_{AA}$ and $\Delta E_{AB}$ monotonously increase, shown in Figure 6a.

$\Delta E_{AA-AB-FM}$ also changes with films' thickness, shown in Figure 6b. When N equals to 2, 3, 4, the corresponding $\Delta E_{AA-AB-FM}$ is 0.042, 0.048, and 0.051 eV, respectively. The positive $\Delta E_{AA-AB-FM}$ means 111-AA-SI-N is higher than 111-AB-SI-N in energy. Most interesting, as N is increased to 5, $\Delta E_{AA-AB-FM}$ becomes -0.084 eV, which means 111-AA-SI-5 is more stable than 111-AB-SI-5, shown in Figure 6b. As film becomes much thicker, the corresponding $\Delta E_{AA-AB-FM}$ monotonously increases to -0.133 ($N = 6$), -0.14 ($N = 7$), -0.148 ($N = 8$), -0.187 ($N = 9$) and -0.239 eV ($N = 10$), respectively. It can be concluded that 111-AA-SI-N are more stable than 111-AB-SI-N stacking patterns, as the film becomes thicker ($N \geq 5$).

The band structures of 111-AA-SI-N and 111-AB-SI-N are also investigated, shown in Figure 6 c, d, respectively. As $N = 3$, both spin-α and spin-β electrons are conducting, which means 111-AA-SI-3 and 111-AB-SI-3 transform from HM ($N = 2$) into normal spin-polarized metal with FM order, shown in Figure 6c. And PDOS are calculated, shown in Figure S12 a, b. As film becomes thicker, the states at Fermi-level also increase, shown in Figure 6 c, d. Taking $N = 4, 6, 8, 10$ as example, the $DOS_F$ of spin-β electrons are 0.452 (0.985), 0.598 (1.726), 0.965 (2.156) 1.215 (2.981) states/eV for 111-AA-SI-N and 111-AB-SI-N, respectively. Therefore, 111-AA-SI-N and 111-AB-SI-N ($N \geq 3$) have changed into normal spin-polarized metal with FM order, shown in Figure 6 c, d, respectively. However, PDOS of 111-AA-SI-N and 111-AB-SI-N ($N = 3, 4$) are also calculated by HSE06, shown in Figure S13 a-d. They are all HM with FM order independent of stacking patterns.

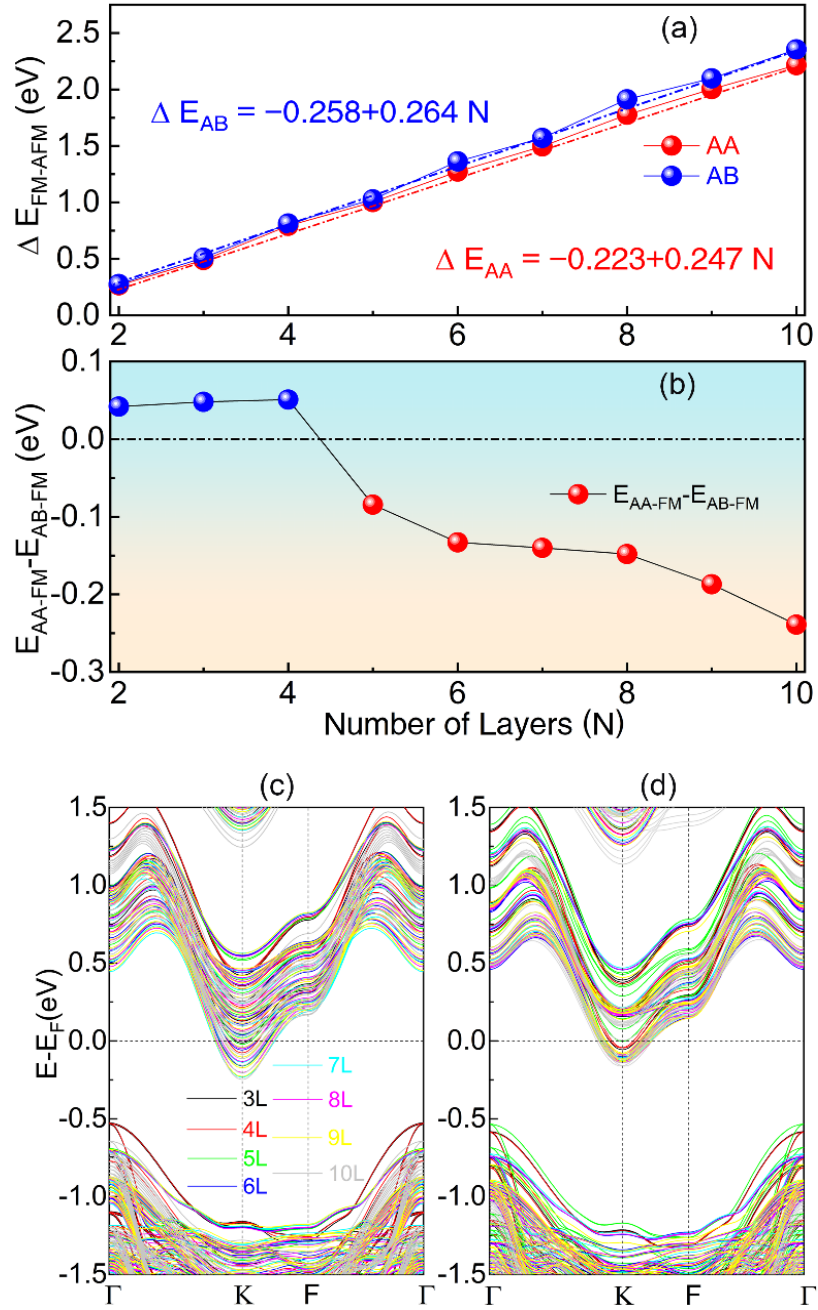

**Figure 6.** (a) The $\Delta E$ and (b) $\Delta E_{AA-AB-FM}$ change with layers. The spin-polarized band structures of (c) 111-AA-SI-N and (d) 111-AB-SI-N (N = 3-10), respectively. The black, red, green, blue, cyan, pink, yellow, grey represent N equals to 3, 4, 5, 6, 7, 8, 9, and 10, respectively. The Fermi-level is set 0 eV.

**3.6. Dynamical and Thermal Stability.** The dynamical stability of SI-CGT is confirmed via phonon dispersion curves and phonon DOS, which show no obvious imaginary phonon modes. The highest vibration frequency is 8.020 and 8.018 THZ for 111-AA-SI-2 and 111-AB-SI-2, which is lower than CGT (8.364 THZ),[27] shown in Figure 7 a, b, S14. From Figure 7b, we can find that Te atoms make main contribution to the low frequency ($0 < \varepsilon < 3.03$ THZ). On the contrary, the high frequency ($6.3 < \varepsilon < 8.06$ THZ) is mainly contributed by Ge atoms, and Cr atoms make main contribution to the middle frequency ($3.03 < \varepsilon < 6.3$ THZ). 111-AA-SI-2 and 111-AB-SI-2 show the similar trend, in Figure 7 a, b.

The thermal stability of SI-CGT is evaluated with AIMD, and the corresponding results are shown in Figure 7 c-f. 111-AA(AB)-SI-N ($N = 2, 3$) are simulated with NVT at 300 K, shown in Figure 7 c, d, e, f, respectively. Total energies vibrate around -386.023, -386.045, -592.876, -593.045 eV, respectively. Meanwhile, both 111-AA-SI-2 and 111-AB-SI-2 have not obvious structure destruction in snapshots at 300 K, shown in Figure S15, 16, respectively. The corresponding magnetic moment are about 72, 72, 120, 120 $\mu_B$, respectively, which implies the geometry and ferromagnetic order are also stable at 500 K, shown in Figure S17, and more detail could be found in the supporting information. Additionally, 111-AA-SI-4 and 111-AB-SI-4 are fabricated, and geometry and magnetic moment are stable at room temperature, shown in Figure S18. More results and discussion could be found in the supporting information.

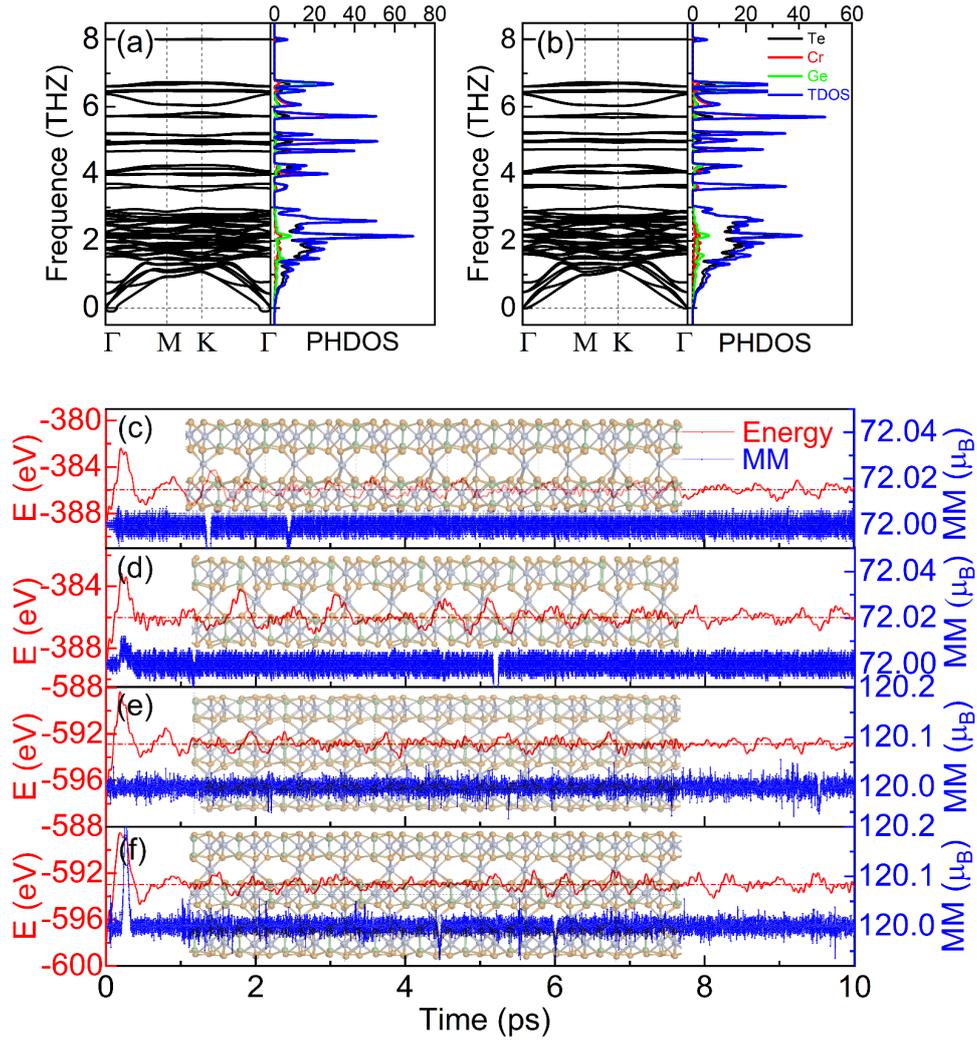

**Figure 7.** The phonon band structures and density state of CGT- (a) 111-AA-SI-2, (b) 111-AB-SI-2. The energy and magnetic moment of (c) 111-AA-SI-2, and (d) 11-AB-SI-2 change with time at 300 K. The black, red, green and blue represent Te, Cr, Ge and total phonon density of the states. The inset shows the snap shots in the simulation. (e) The energy and magnetic moment of (e) 111-AA-SI-3 and (f) 111-AB-SI-3 change with time at 300 K, respectively.

**3.7. Magnetocrystalline Anisotropy.** In order to clarify atomic orbital contribution to the MAE, the tight-binding and second-order perturbation theory[27] are used. MAE of each atom could be evaluated, based on canonical formulation. The MAE could be written in the following equation:

$$MAE_i = \left[ \int E_f (E - E_F)[n_i^{[100]}(E) - n_i^{[001]}(E)] \right] \quad (6)$$

where $MAE_i$ represents MAE contributed by the $i$th atom. $n_i^{[100]}(E)$ and $n_i^{[001]}(E)$ represent DOS of the $i$th atom with EA along [100] and [001] directions, respectively. In addition, MAE could be rewritten as sum of $MAE_i$, shown in the following equation:

$$MAE_{tot} = \sum_i MAE_i \quad (7)$$

Based on the second-order perturbation theory,[66] MAE could be evaluated by the sum of the following terms:

$$\Delta E^{--} = E_x^{--} - E_z^{--} = \xi^2 \sum_{o^+, u^-} (|<o^-|L_z|u^-)|^2 - |<o^-|L_x|u^->|^2)/(E_u^- - E_o^-) \quad (8)$$

$$\Delta E^{-+} = E_x^{+-} - E_z^{+-} = \xi^2 \sum_{o^+, u^-} (|<o^+|L_z|u^-|^2 - |<o^+|L_x|u^->|^2)/(E_u^- - E_o^-) \quad (9)$$

where + and − represent spin-α and spin-β electrons states, and $\xi$, $L_x$, $L_z$ are SOC constant, angular momentum operators along [100] and [001] directions, respectively. *u*, and *o* represent unoccupied and occupied states, respectively. $E_o$, $E_u$ represent energies of occupied and unoccupied states, respectively. According to the above equations, MAE mainly come from contribution of spin-orbital matrix elements and energy difference. The matrix element differences $<o^-|L_z|u^-\rangle|^2 - |<o^-|L_x|u^-\rangle|^2$ and $|<o^+|L_z|u^-\rangle|^2 - |<o^+|L_x|u^-\rangle|^2$ for *p* and *d* orbitals are calculated, shown in Table 1 and Table 2, respectively. In order to analyze change of MAE with stacking orders, the atom-orbital-resolved MAE is calculated, shown in Figure 8 a-l, S19.

It can be concluded that MAE of 111-AA(AB)-2 without $Cr_{SI}$ atoms and 111-AA(AB)-SI-2 mainly come from Cr and Te atoms' contribution, shown in Figure 8 a-d, i-l. However, it partially comes from Ge's contribution, shown in Figure 8 e-h, respectively. The total MAEs of 111-AA(AB)-2 are 1.80 (1.91) mev/.f.u., and Ge atoms contribute 0.015 (0.019) meV, while Cr atoms contribute 0.698 (0.698) meV, shown in Figure 8 a, b, e, f. However, Te atoms of 111-AA(AB)-2 contribute 1.08 and 1.32 meV, respectively. The hybridization between Te's $p_y$ and $p_z$ orbitals of 111-AA-2, which corresponds to the matrix differences 1 for *p* orbitals, intends PMA (6.86 meV), shown in Table 1. However, the hybridization between Te's $p_y$ and $p_x$ orbitals, which corresponds to the matrix differences −1 for *p* orbitals, intends IMA (-5.6 meV).

**Table 1.** The matrix differences for $p$ orbitals between EA along [001] and [100] directions in eq 12 and eq 13.

| | $o^+$ | | | $o^-$ | | |
|---|---|---|---|---|---|---|
| $u^-$ | $p_y$ | $p_z$ | $p_x$ | $p_y$ | $p_z$ | $p_x$ |
| $p_y$ | 0 | 1 | -1 | 0 | -1 | 1 |
| $p_z$ | 1 | 0 | 0 | -1 | 0 | 0 |
| $p_x$ | -1 | 0 | 0 | 1 | 0 | 0 |

When Cr$_{SI}$ atoms are covalently bonding with Te atoms, total MAE of 111-AA-SI-2 is decreased to -0.160 meV/.f.u., and Ge atoms contribute -0.082 meV, while Cr atoms contribute 1.011 meV, shown in Figure 8 k, c. The hybridization between Cr's $d_{yz}$ and $d_{z^2}$, $d_{xz}$ and $d_{xy}$, $d_{x^2-y^2}$ and $d_{yz}$ orbitals contribute 1.359, 0.75, 0.75 meV to MAE, which correspond to the matrix differences 3, 4 and 1 for $d$ orbitals, shown in Table 2. And these $d$ orbitals hybridization favors PMA. The hybridization between Cr atoms' $d_{xy}$ and $d_{x^2-y^2}$, $d_{xz}$ and $d_{yz}$ orbitals contribute -1.174, -0.664 meV to MAE, which correspond to the matrix differences $-4$, and $-1$ for $d$ orbitals, respectively. And it means that this hybridization favors IMA. In a word, the $d$ orbitals make positive contribution to MAE, which is similar with CGT bilayers (Figure 8 a, b). In addition, Te atoms contribute -1.086 meV to the total MAE. The hybridization between Te's $p_y$ and $p_z$ orbitals intends PMA (2.240 meV). However, the hybridization between Te atoms' $p_x$ and $p_y$ orbitals, $p_x$ and $p_z$ contribute -3.057, -0.227 meV, intending IMA.

**Table 2.** The matrix differences for $d$ orbitals between magnetization along [001] and [100] directions in eq 14 and eq 15.

| $u^-$ | $o^+$ | | | | | $o^-$ | | | | |
|---|---|---|---|---|---|---|---|---|---|---|
| | $d_{xy}$ | $d_{yz}$ | $d_{z^2}$ | $d_{xz}$ | $d_{x^2-y^2}$ | $d_{xy}$ | $d_{yz}$ | $d_{z^2}$ | $d_{xz}$ | $d_{x^2-y^2}$ |
| $d_{xy}$ | 0 | 0 | 0 | 1 | -4 | 0 | 0 | 0 | -1 | 4 |
| $d_{yz}$ | 0 | 0 | 3 | -1 | 1 | 0 | 0 | -3 | 1 | -1 |
| $d_{z^2}$ | 0 | 3 | 0 | 0 | 0 | 0 | -3 | 0 | 0 | 0 |
| $d_{xz}$ | 1 | -1 | 0 | 0 | 0 | -1 | 1 | 0 | 0 | 0 |
| $d_{x^2-y^2}$ | -4 | 1 | 0 | 0 | 0 | 4 | -1 | 0 | 0 | 0 |

For 111-AB-SI-2, the total MAE is 0.428 meV/.f.u., Ge atoms contribute -0.082 meV to total MAE, shown in Figure 8e. And Cr atoms contribute 0.744 meV to MAE, shown in Figure 8d. The hybridization between Cr's $d_{yz}$ and $d_{z^2}$, $d_{xz}$ and $d_{xy}$, $d_{x^2-y^2}$ and $d_{yz}$ orbitals intends PMA (1.361, 0.61, 0.61 meV), shown in Figure 8d. However, hybridization between Cr atoms' $d_{xy}$ and $d_{x^2-y^2}$ orbitals, $d_{xz}$ and $d_{yz}$ intends IMA (-1.151, -0.681 meV). Compared with 111-AA-SI-2, Te atoms in 111-AB-SI-2 contribute -1.086 meV to total MAE. And the hybridization between Te's $p_y$ and $p_z$ orbitals contributes 1.704 meV, shown in Figure 8l. However, the hybridization between Te atoms' $p_x$ and $p_y$ orbitals, $p_x$ and $p_z$ contributes -2.692, -0.098 meV to IMA. In 111-AB-2, the hybridization between Te's $p_y$ and $p_z$ orbitals contributes 6.98 meV to PMA, while hybridization between Te's $p_y$ and $p_x$ orbitals contributes -5.5 meV to PMA, shown in Figure 8j. Compared with 111-AB-2, 111-AB-SI-2 shows different trend. In a word, the Te atoms' contribution to MAE is obviously reduced, as the hybridization between Te's $p_y$ and $p_z$ orbitals, $p_y$ and $p_x$ orbitals is obviously weakened, when Cr$_{SI}$ atoms are introduced. Therefore, the EA is switched from original [001] to [100] direction.

MAE of 111-AA-SI-2 is 0.267 meV larger than 111-AB-SI-2, shown in Figure 8. The Te, Ge make the same contribution to the total MAE. The difference come from the Cr atoms' contribution, shown in Figure 8 c, d. The different stacking orders affect the hybridization between Cr atoms' $d_{xz}$ and $d_{xy}$, $d_{xy}$ and $d_{x^2-y^2}$ orbitals, which contribute 0.75, 0.75 meV (111-AA-SI-2), and 0.61, 0.61 meV (111-AB-2), shown in Figure 3, 8 c, d, respectively.

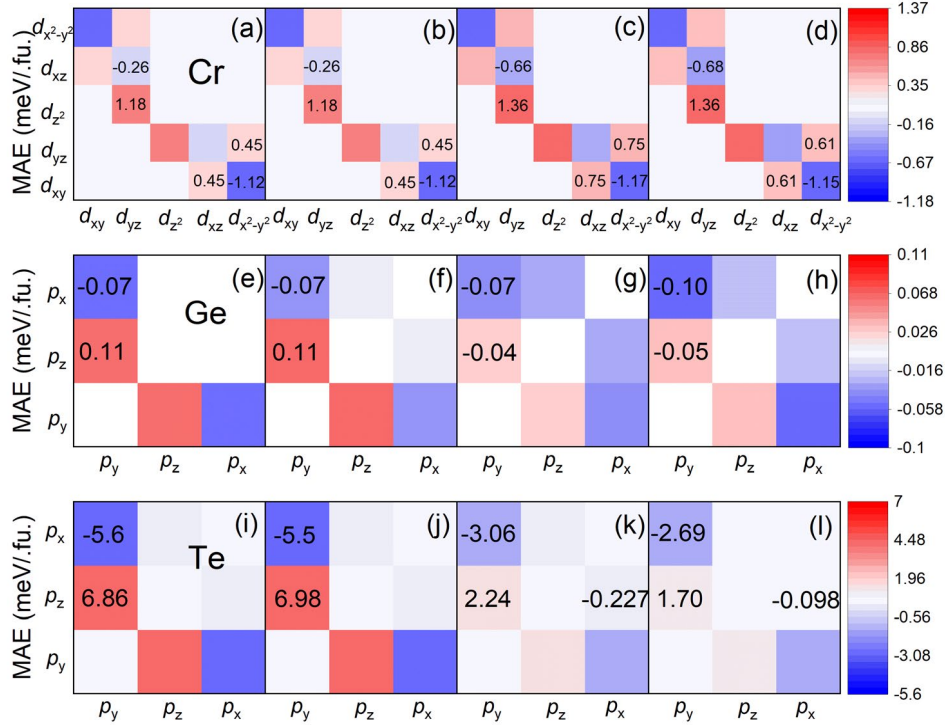

**Figure 8.** (a, b, c, d) Cr's *d*-orbital, (e, f, g, h) Ge's *p*-orbital, and (i, j, k, l) Te's *p*-resolved MAE of AA-CGT bilayer, AB-CGT bilayer, CGT-111-AA-SI-2 and CGT-111-AB-SI-2 bilayers, respectively.

**3.8. Migration barrier of Cr$_{SI}$ atoms.** The migration barrier for the intercalated Cr atom to move within the interstitial vdW gap in 111-AA(AB)-SI-2 is calculated using the NEB method. Six images are inserted between the beginning and ending point, shown in Figure 9a. The migrated barriers of Cr$_{SI}$ atoms in 111-AA(AB)-SI-2 are 1.305 and 1.193 eV, shown in Figure 9 a, b, respectively. The Cr$_{SI}$ atom of SI-CGT migrates from hollow site (the hollow sites of top and bottom layers) to the nearby hollow site, shown in Figure 9 a, b. The migrated barriers of Cr$_{SI}$ atom are high, which implies that Cr$_{SI}$ atoms are uneasy to shift from hollow site to the nearby hollow site. The barrier of 111-AA-SI-2 is higher than 111-AB-SI-2, which implies that the migration of $Cr_{SI}$ atom in 111-AA-SI-2 need more energy to overcome the barrier.

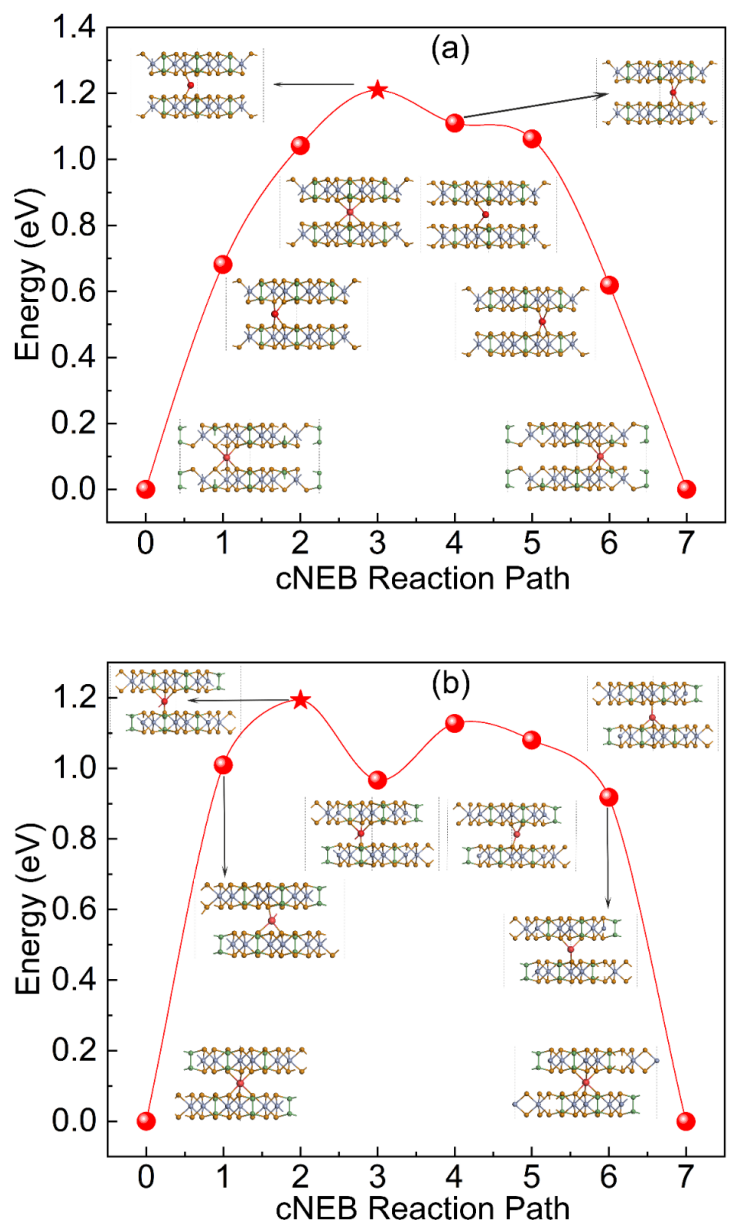

**Figure 9.** The migration barrier of $Cr_{SI}$ atom of (a) 111-AA-SI-2 and (b) 111-AB-SI-2, respectively. The star represents the transition state structures.

## 3. CONCLUSIONS

In summary, we have systemically investigated magnetic and electronic properties of SI-CGT. We have found interlayer exchange interaction in the self-intercalated system. The SI-CGT bilayer is HM with FM order, independent of stacking orders. The phonon spectrum and AIMD confirm structural and magnetic stability of SI-CGT bilayers. The EAs of SI-CGT bilayers intend IMA, with MAE of -0.160 (AA), and -0.428 (AB) meV/f.u., respectively. The orbital-resolved MAEs show that Te's contribution dominates MAE of SI-CGT bilayers. The switch of EA origins hybridization between Te's $p_x$ and $p_y$, $p_y$ and $p_z$ is obviously weakened. SI-CGTs are HM with FM order, independent of $Cr_{SI}$ atoms' concentration. When films of SI-CGT become thicker ($N \geq 3$), they transform into normal spin-polarized metal. The migrated barriers of $Cr_{SI}$ atoms in 111-AA(AB)-SI-2 are 1.305 and 1.193 eV, respectively. Our findings provide new strategy to control magnetoelectric properties of vdW magnetic materials.

**Supporting Information**

Information on materials, U test on MAE, k-mesh test, differential charge density of SI-CGT, change of $\Delta E$ with magnetic orders, band structure of different orders, magnetoelectric properties calculated by DFT-D3, PDOS calculated by HSE06, band structure with SOC, magnetoelectric properties of $\sqrt{3}\times\sqrt{3}\times1$ cell, magnetic anisotropy properties of CGT bilayers, PDOS of 111-AA(AB)-3(6) calculated by PBE+U, PDOS calculated by HSE06, phonon spectrum simulation, snap shots of AIMD, AIMD at 500 K, AIMD of four layers, projected MAE of SI-CGT bilayers.

**Author Information**

Corresponding Author

*E-mail: zyguan@sdu.edu.cn; Tel: +86-0531-88363179; Fax: +86-0531-88363179

**Acknowledgements**

We thank Prof. Wenhui Duan and Xingxing Li for discussion of evaluation of MAE. We thank Doctor Rui Li for discussion of orbital-resolved MAE. We thank doctor Yang Li for discussion of LDA+U. This work was supported by the financial support from the Natural Science Foundation of China (Grant No. 11904203), and the Fundamental Research Funds of Shandong University (Grant No. 2019GN065). The authors are grateful to Beijing PARATERA Tech Corp., Ltd. for the computation resource in the National Supercomputer Center of Guangzhou. The scientific calculations in this paper have been done on the HPC Cloud Platform of Shandong University. The authors also acknowledge the Beijing Super Cloud Computing Center (BSCC), Shanghai Supercomputer Center and Tencent Quantum Laboratory for providing HPC resources.

**Conflict of Interest:** The authors declare no competing financial interest.